\pdfoutput=1
\documentclass[11pt]{article}

\usepackage[a4paper,margin=1in]{geometry}
\usepackage{amsmath,amssymb,amsthm}
\usepackage{graphicx}
\usepackage{booktabs}
\usepackage{algorithm}
\usepackage{algpseudocode}
\usepackage[table]{xcolor}
\usepackage{caption}
\usepackage{subcaption}
\usepackage{tikz}
\usetikzlibrary{positioning,arrows.meta,fit,backgrounds,calc}
\usepackage{enumitem}
\usepackage{amsfonts}
\usepackage[colorlinks=true,linkcolor=blue!55!black,citecolor=green!45!black,
            urlcolor=blue!55!black]{hyperref}

\newtheorem{problem}{Problem}
\DeclareMathOperator*{\argmax}{arg\,max}

\newcommand{\E}{\mathbb{E}}

\newcommand{\comp}{\mathbf{w}}
\newcommand{\model}{\textsc{TFR-GNN}}
\newcommand{\heft}{\textsc{HEFT}}
\newcommand{\peft}{\textsc{PEFT}}
\newcommand{\cpop}{\textsc{CPOP}}
\newcommand{\rheft}{\textsc{R-HEFT}}
\newcommand{\ftheft}{\textsc{FT-HEFT}}

\title{\vspace{-1.2cm}\textbf{\model{}: Topology- and Fault-Aware Graph Neural
Scheduling for Heterogeneous Distributed Computing Systems}}

\author{
\normalsize Shiyu Yang\,$^{1}$ \quad Ziyang Zeng\,$^{2}$ \quad Jie-Si Yang\,$^{3}$
\\[4pt]
\normalsize $^{1}$University of California, Los Angeles \quad
$^{2}$New York University \quad
$^{3}$University of Utah
\\[-2pt]
}
\date{}

\begin{document}
\maketitle
\vspace{-0.8cm}

\begin{abstract}
\noindent
Scheduling workflow directed acyclic graphs (DAGs) on heterogeneous distributed
systems is a classical NP-hard problem, and the list-scheduling heuristic
\heft{} remains the de-facto standard because of its low complexity and strong
makespan. In real deployments, however, machines fail: commodity and
``spot''/pre-emptible nodes are far less reliable than dedicated ones, and a
makespan-optimal but reliability-agnostic placement can be dramatically slowed
by node failures. We show empirically, on real workflow structures from the
WfCommons/Pegasus corpus, that \emph{no single fixed heuristic} is best across
the joint space of failure intensity and cluster load: with no failures \heft{}
is optimal, whereas under failures a reliability-aware placement can reduce the
expected makespan by up to $52\%$ when spare capacity exists. Motivated by this,
we present \model{}, a graph neural scheduler that combines (i)~bidirectional
\emph{dependency attention} over the task DAG, (ii)~\emph{topology attention}
over the (bandwidth-weighted) machine graph, and (iii)~a task$\leftrightarrow$
machine \emph{cross-attention} placement head augmented with a
failure-\emph{gated} reliability tilt and an optional replication gate. Because
policy-gradient exploration fails to discover the required coordinated placement
on a modest compute budget, we train \model{} by \emph{distilling a
best-of-portfolio fault-tolerant oracle} into a single one-shot policy. On real
workflows and a bimodal-reliability cluster model, \model{} (a)~matches \heft{}
\emph{exactly} when there are no failures (a built-in adaptivity property),
(b)~reduces the expected makespan under failures by $14.8\%$ on average
(up to $47\%$) over \heft{}, (c)~beats a fixed reliability-aware baseline
(\rheft{}) by $\sim\!11\%$ and \emph{matches a per-scenario hindsight oracle}
($0.998\times$) as a single policy without any deployment-time tuning, and
(d)~generalises to unseen applications and to workflows an order of magnitude
larger than those seen in training, while producing schedules in well under a
second for graphs of nearly $5{,}000$ tasks. All results are produced by a
verified event-level simulator on real workflow data; no experimental numbers
are synthetic.

\vspace{4pt}
\end{abstract}

\section{Introduction}\label{sec:intro}

Modern scientific and data-analytics applications are commonly expressed as
\emph{workflows}: directed acyclic graphs (DAGs) whose nodes are compute tasks
and whose edges encode data dependencies. Executing such a workflow on a
distributed platform requires a \emph{schedule}---an ordering of the tasks and
an assignment of each task to a machine---that minimises the overall completion
time (\emph{makespan}). When machines are \emph{heterogeneous} (different speeds
and task affinities) and communication is non-uniform, this is the classical
NP-hard heterogeneous DAG-scheduling problem~\cite{topcuoglu2002,braun2001}.

For two decades the Heterogeneous Earliest-Finish-Time heuristic
(\heft{})~\cite{topcuoglu2002} has been the de-facto standard: it ranks tasks by
an \emph{upward rank} along the critical path and greedily places each ready
task on the machine giving the earliest finish time. \heft{} is $O(v^2 p)$ for
$v$ tasks and $p$ processors, and its simplicity and competitive makespan have
made it the baseline against which virtually all subsequent
heuristics---\cpop{}~\cite{topcuoglu2002}, \peft{}~\cite{arabnejad2014} and
many others---are measured.

\paragraph{The fault-tolerance gap.}
\heft{} and its descendants optimise makespan on an idealised, \emph{failure-free}
platform. Real platforms are not failure-free. Commodity clusters experience
node crashes, and cloud ``spot''/pre-emptible instances---attractive because
they are cheap---can be reclaimed or fail an order of magnitude more often than
dedicated ``on-demand'' nodes. Crucially, reliability is \emph{not} aligned with
speed: a fast node may well be a volatile one. A scheduler that greedily chases
the earliest finish time will therefore pile critical work onto fast-but-volatile
machines and be repeatedly set back by failures, restarts and lost work. The
right response---placing critical tasks on reliable machines, and occasionally
replicating them---trades a little raw speed for a large reduction in the
\emph{expected} makespan under failures.

\paragraph{No single heuristic is best everywhere.}
A central empirical observation of this paper (Section~\ref{sec:results}) is that
the best scheduling policy depends jointly on the \emph{failure intensity} and
the \emph{cluster load}:
\begin{itemize}[leftmargin=1.3em,itemsep=1pt,topsep=2pt]
\item with no failures, \heft{} is optimal and any fault-tolerance machinery is
pure overhead;
\item under failures \emph{with spare capacity}, steering critical tasks away
from volatile nodes (reliability-aware placement) reduces the expected makespan
by up to $52\%$;
\item under failures \emph{with tight capacity}, the gains shrink and selective
task replication occasionally helps.
\end{itemize}
A practitioner deploying a fixed heuristic must therefore guess a reliability
weight and a replication budget \emph{per scenario}---exactly the kind of manual
tuning that a learned, \emph{adaptive} policy should eliminate.

\paragraph{Why learning, and why not na\"ive reinforcement learning.}
A scheduler that observes the workflow structure, the per-machine reliability,
the failure intensity and the current load could in principle produce the right
schedule for \emph{every} regime with a single forward pass. Graph neural
networks (GNNs) with attention are a natural fit for the DAG-plus-cluster input.
However, we find (Section~\ref{sec:method-train}) that training such a policy
\emph{from scratch} with policy gradients does not work on a modest
(single-CPU) budget: \heft{} is a strong local optimum, almost every local
perturbation is worse, and even a leave-one-out (RLOO) estimator---which removes
the systematic drift of self-critical REINFORCE---fails to \emph{discover} the
coordinated, reliability-aware placement that we know exists and is worth up to
$52\%$. We therefore adopt a reliable alternative: we build a
\emph{best-of-portfolio fault-tolerant oracle} per scenario and \emph{distil} it
into a single one-shot GNN policy.

\paragraph{Contributions.} This paper makes the following contributions.
\begin{enumerate}[leftmargin=1.5em,itemsep=2pt,topsep=2pt]
\item \textbf{A topology- and fault-aware graph neural scheduler} (\model{})
that couples bidirectional \emph{dependency attention} over the task DAG,
\emph{topology attention} over the bandwidth-weighted machine graph, and a
task$\leftrightarrow$machine \emph{cross-attention} placement head. A
\emph{failure-activity gate} makes all fault-tolerance mechanisms vanish when
there are no failures, so the policy provably reduces to min-EFT/\heft{} at zero
failure intensity---an explicit adaptivity guarantee
(Section~\ref{sec:method}).
\item \textbf{A distillation training recipe} that turns a portfolio of
classical and fault-tolerant heuristics (\heft{}, reliability-aware \rheft{} at
several strengths, and replication-based \ftheft{} at several budgets) into a
single adaptive policy, after showing that direct policy-gradient learning is
inadequate here. The learned policy reproduces best-of-portfolio behaviour
across the load$\times$failure grid \emph{without} per-scenario tuning.
\item \textbf{A thorough empirical study on real workflows} from
WfCommons/Pegasus~\cite{wfcommons,pegasus} using a verified event-level
simulator and a realistic \emph{bimodal} reliability model. \model{} matches
\heft{} exactly without failures, reduces the expected makespan by $14.8\%$ on
average (up to $47\%$) under failures, beats a fixed reliability-aware baseline
by $\sim\!11\%$, and \emph{matches a per-scenario hindsight oracle}. We further
report ablations, sensitivity analyses, generalisation to unseen applications
and unseen (up to $10\times$ larger) sizes, and scalability. Every number is
computed by simulation on real workflow data.
\end{enumerate}

The remainder of the paper reviews related work
(Section~\ref{sec:related}), formalises the problem
(Section~\ref{sec:problem}), describes \model{} and its training
(Section~\ref{sec:method}), details the experimental setup
(Section~\ref{sec:setup}), presents results
(Section~\ref{sec:results}), discusses limitations
(Section~\ref{sec:discussion}) and concludes
(Section~\ref{sec:conclusion}).

\section{Related Work}\label{sec:related}

\paragraph{List scheduling for heterogeneous DAGs.}
Topcuo\u{g}lu et al.~\cite{topcuoglu2002} introduced \heft{} and \cpop{}; \heft{}
ranks tasks by upward rank and greedily minimises earliest finish time, while
\cpop{} prioritises the critical path. Arabnejad and
Barbosa~\cite{arabnejad2014} proposed \peft{}, which adds a look-ahead through
an \emph{optimistic cost table} (OCT), improving processor selection at the same
$O(v^2p)$ complexity. For \emph{independent} tasks, Braun et
al.~\cite{braun2001} compared eleven static heuristics and found Min-Min and
Max-Min to be simple yet effective; we include both as baselines. These methods
target failure-free makespan and do not model reliability.

\paragraph{Fault-tolerant and reliability-aware scheduling.}
Fault tolerance in DAG scheduling is classically addressed by
\emph{replication} (running copies of a task on distinct machines) and by
\emph{reliability-aware placement} (biasing critical tasks towards dependable
machines). Replication improves robustness at the cost of redundant computation
and contention, whereas reliability-aware placement trades a little speed for
fewer restarts. A recurring difficulty is that the right amount of each depends
on the operating regime. Our baselines \rheft{} and \ftheft{}
(Section~\ref{sec:setup}) instantiate these two ideas as reliability-tilted and
replication-augmented variants of \heft{}; \model{} learns to combine and
modulate them automatically.

\paragraph{Learning-based scheduling.}
Learning to schedule has attracted significant attention. Decima~\cite{decima}
uses a graph neural network trained with reinforcement learning to schedule
data-processing DAGs on a cluster, demonstrating that GNN policies can match or
exceed hand-tuned heuristics. Subsequent work has explored reinforcement
learning and imitation learning for DAG and job scheduling. Our work differs in
three ways: (i)~we target \emph{fault tolerance} on heterogeneous machines with
non-uniform reliability, not throughput on homogeneous clusters; (ii)~we show
that na\"ive policy gradients are inadequate in our regime and instead
\emph{distil} a fault-tolerant oracle; and (iii)~we build an explicit adaptivity
guarantee (reduction to \heft{} without failures) into the architecture.

\paragraph{Graph neural networks and attention.}
Our encoder builds on graph neural networks~\cite{gcn,graphsage} and graph
attention~\cite{gat}, and on the (dense) scaled dot-product attention of the
Transformer~\cite{transformer}, which we apply over the machine graph with a
bandwidth bias. The distillation objective follows the standard idea of training
a compact student to imitate a stronger teacher; here the teacher is a portfolio
of scheduling heuristics rather than a single large model. For completeness we
also relate our (unsuccessful) reinforcement-learning attempts to
REINFORCE~\cite{reinforce}, its self-critical variant~\cite{scst}, the
leave-one-out estimator (RLOO)~\cite{rloo} and PPO~\cite{ppo}.

\section{Problem Formulation}\label{sec:problem}

\subsection{Workflow and platform model}
A workflow is a DAG $G=(V,E)$ with $n=|V|$ tasks. Each task $i$ has a
computational \emph{work} $w_i>0$ and a type $\tau_i$; each edge $(i,j)\in E$
carries a data volume $d_{ij}\ge 0$ that must be transferred if $i$ and $j$ run
on different machines. The platform is a set $M=\{1,\dots,p\}$ of heterogeneous
machines. Machine $m$ has a relative speed $s_m$ and a per-type affinity
$e_{\tau m}>0$; the \emph{computation cost} of task $i$ on machine $m$ is
\begin{equation}
\comp_{im} \;=\; w_i \, e_{\tau_i m}\,/\,s_m .
\label{eq:comp}
\end{equation}
Machines are grouped into racks; the \emph{communication cost} of edge $(i,j)$
placed on machines $(a,b)$ is
\begin{equation}
c_{ij}(a,b)=
\begin{cases}
0, & a=b,\\[2pt]
d_{ij}/B(a,b) + L, & a\neq b,
\end{cases}
\label{eq:comm}
\end{equation}
where $B(a,b)$ is the bandwidth (high intra-rack, low inter-rack) and $L$ a
latency. This is the standard related-machines model underlying
\heft{}~\cite{topcuoglu2002,arabnejad2014}.

\subsection{Bimodal reliability (fault) model}\label{sec:faultmodel}
Each machine $m$ fails as a Poisson process with rate $\lambda_m$; a failure is
followed by a repair time drawn from a log-normal distribution with mean
$\mu_r$. A task running on $m$ when it fails is lost and \emph{restarted} after
repair (its partial progress is wasted). To model the reality of mixed
dedicated/spot fleets, reliability is \emph{bimodal and independent of speed}: a
fraction $\rho$ of machines are \emph{volatile} (short mean time between
failures, MTBF $\in[60,250]\,$s) and the remainder are \emph{reliable}
(MTBF $\in[3\!\times\!10^3,3\!\times\!10^4]\,$s); speeds $s_m$ are drawn
independently, so fast machines are as likely to be volatile as reliable. A
stress parameter $s\ge 0$ (the failure-intensity \emph{scale}) multiplies every
$\lambda_m$, so $s{=}0$ is failure-free and larger $s$ is harsher. The
steady-state \emph{availability} of machine $m$ is
$a_m=\mathrm{MTBF}_m/(\mathrm{MTBF}_m+\mu_r)\in[0,1]$, and its expected downtime
fraction at scale $s$ is
$\delta_m(s)=\lambda_m s\,\mu_r/(1+\lambda_m s\,\mu_r)$, with
$\delta_m(0)=0$.

\subsection{Schedule and objective}
A schedule $\pi$ specifies, for each task, an ordering priority and one or more
machines (a \emph{primary} and, optionally, a \emph{replica}); execution follows
non-insertion list scheduling with the failure/restart semantics above. Given a
sampled failure trace $\omega$ (the down-intervals of every machine over the
horizon), the simulator returns the realised makespan
$C_{\max}(\pi,\omega)$, the \emph{wasted} work (progress lost to restarts) and
the \emph{redundant} work (extra computation from replicas). We evaluate the
\emph{expected} makespan over the failure distribution,
$\E_\omega[C_{\max}(\pi,\omega)]$, estimated by Monte-Carlo over $K$ traces.

\begin{problem}[Fault-tolerant heterogeneous DAG scheduling]
Given $G$, a platform with computation/communication costs
\eqref{eq:comp}--\eqref{eq:comm}, a failure model with intensity $s$
(Section~\ref{sec:faultmodel}) and an optional replication budget $\beta$
(maximum fraction of tasks that may be replicated), find a schedule $\pi$
minimising $\E_\omega[C_{\max}(\pi,\omega)]$.
\end{problem}

At $s=0$ this reduces to classical makespan minimisation, for which \heft{} is
the standard strong heuristic; for $s>0$ the optimal policy must weigh speed
against reliability, and the balance depends on $s$ and on the available spare
capacity (Section~\ref{sec:results}).

\section{The \model{} Scheduler}\label{sec:method}

\subsection{Overview}
\begin{figure}[t]
\centering
\begin{tikzpicture}[
  font=\small,
  box/.style={draw,rounded corners=2pt,align=center,inner sep=4pt,
              minimum height=8mm,fill=blue!5},
  fhead/.style={draw,rounded corners=2pt,align=center,inner sep=4pt,
              minimum height=8mm,fill=red!6},
  io/.style={align=center,inner sep=2pt},
  ar/.style={-{Latex[length=2mm]},thick}
]
\node[io] (dag) at (0,1.1) {task DAG\\{\footnotesize$(w_i,\tau_i,d_{ij})$}};
\node[io] (mach) at (0,-1.1) {machines\\{\footnotesize$(s_m,a_m,\delta_m,\mathrm{rack})$}};
\node[box] (nenc) at (2.5,1.1) {node\\encoder};
\node[box] (menc) at (2.5,-1.1) {machine\\encoder};
\node[box] (dep) at (5.0,1.1) {dependency\\attention\\{\footnotesize(bi-dir GAT)}};
\node[box] (topo) at (5.0,-1.1) {topology\\attention\\{\footnotesize(BW-biased)}};
\node[box] (cross) at (7.7,0) {cross\\attention};
\node[fhead] (fault) at (10.3,0.9) {fault-aware\\placement tilt\\{\footnotesize(gated by $a(s)$)}};
\node[fhead] (rep) at (10.3,-0.9) {replication\\gate};
\node[box] (dec) at (12.7,0) {guided\\decode\\{\footnotesize(Alg.~1)}};
\node[io] (out) at (14.3,0) {schedule\\$\pi$};

\draw[ar](dag)--(nenc); \draw[ar](mach)--(menc);
\draw[ar](nenc)--(dep); \draw[ar](menc)--(topo);
\draw[ar](dep)-- (cross.north west); \draw[ar](topo)-- (cross.south west);
\draw[ar](cross)--(fault); \draw[ar](dep.east)  to[out=0,in=150] (rep.west);
\draw[ar](topo.east) to[out=0,in=210] (rep.west);
\draw[ar](fault)--(dec); \draw[ar](rep)--(dec);
\draw[ar](dec)--(out);
\begin{scope}[on background layer]
\node[draw,dashed,rounded corners,fit=(dep)(topo)(cross),
      inner sep=6pt,label=below:{\footnotesize attention encoder}]{};
\end{scope}
\end{tikzpicture}
\caption{\model{} architecture. A single forward pass produces task priorities
and placement scores; a linear-time greedy decode (Algorithm~\ref{alg:decode})
turns them into a schedule. All fault-tolerance mechanisms are gated by the
failure activity $a(s)$, which is zero when there are no failures.}
\label{fig:arch}
\end{figure}
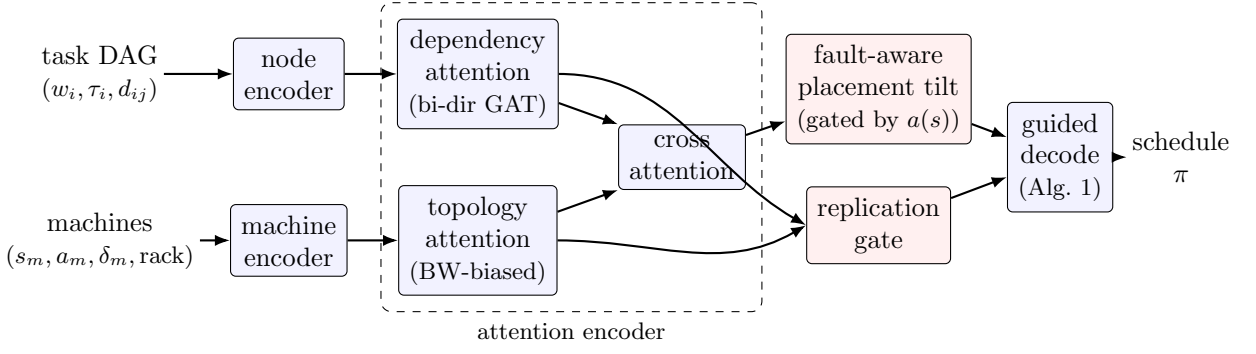

\model{} maps a workflow and a cluster to a schedule with a \emph{single}
forward pass followed by a linear-time greedy decode. Figure~\ref{fig:arch}
sketches the architecture. Node and machine features are encoded and fused with
a broadcast context vector; \emph{dependency attention} refines task embeddings
along the DAG; \emph{topology attention} refines machine embeddings over the
bandwidth graph; a \emph{cross-attention} head produces a task$\leftrightarrow$
machine placement affinity; and a \emph{fault-aware head} adds a
failure-gated reliability tilt and an optional replication gate. The resulting
priorities and placement scores drive a guided list-scheduling decode
(Section~\ref{sec:decode}). The model has only $62{,}870$ parameters and runs on
a single CPU.

\subsection{Features}\label{sec:features}
Each task carries a $9$-dimensional feature vector (standardised log-work,
log-memory, average CPU count, log-output-volume, log in/out degree, normalised
upward and downward ranks, and a \emph{criticality} score
$\kappa_i=(r^{u}_i+r^{d}_i)/\mathrm{CP}\in[0,1]$ relating the task to the
critical path length $\mathrm{CP}$) together with a learned type embedding. Each
machine carries a $5$-dimensional vector (relative and log speed, availability
$a_m$, downtime fraction $\delta_m(s)$, and log mean computation cost) plus a
learned rack embedding. A $6$-dimensional context vector summarises the scenario
($\log n$, $\log p$, load $n/p$, communication-to-computation ratio, the failure
scale $s$, and the replication budget $\beta$) and is broadcast to all
task/machine encoders. Edges carry the standardised log data-volume.

\subsection{Dependency attention}
Task embeddings $h_i$ are refined by $L$ graph-attention
layers~\cite{gat} applied in \emph{both} directions along the DAG (an added
binary feature marks edge direction), so a task can gather information from both
ancestors and descendants---capturing critical-path and communication structure.
Each layer uses multi-head additive attention with edge features, a residual
connection and layer normalisation.

\subsection{Topology attention}
Machine embeddings $g_m$ are refined by dense multi-head scaled dot-product
attention~\cite{transformer} over all machines, with an \emph{additive
log-bandwidth bias} $\log B(a,b)$ on the attention logits so that machines that
communicate cheaply attend to one another. This yields heterogeneity- and
reliability-aware machine representations.

\subsection{Cross-attention placement and the fault-aware head}
\label{sec:fault-head}
A placement affinity is produced by projecting task and machine embeddings into
a common space and taking scaled dot products,
$\mathrm{cross}_{im}=\langle W_t h_i,\,W_m g_m\rangle/\sqrt{d_c}$. The final
placement bias combines this affinity with a reliability tilt:
\begin{equation}
b_{im}\;=\;a(s)\,\theta\,\tanh(\mathrm{cross}_{im})\;-\;\gamma(\text{ctx})\,
\kappa_i\,\delta_m(s),
\label{eq:bias}
\end{equation}
where $\theta=\mathrm{softplus}(\theta_0)$ is a learned cross weight and
$\gamma(\text{ctx})=\mathrm{softplus}(\gamma_0+\mathrm{MLP}(\text{ctx}))>0$ is a
\emph{context-dependent} reliability weight, allowing the tilt strength to adapt
to the failure intensity and load. The key term is the
\emph{failure-activity gate}
\begin{equation}
a(s)=\tanh\!\big(4\,\overline{\delta}(s)\big)\in[0,1),\qquad
\overline{\delta}(s)=\tfrac{1}{p}\textstyle\sum_m \delta_m(s),
\label{eq:gate}
\end{equation}
which is \emph{exactly zero without failures} (since $\delta_m(0)=0$). Both terms
in \eqref{eq:bias} then vanish at $s{=}0$, so the placement score reduces to
$-\,\mathrm{EFT}$ and \model{} degrades \emph{exactly} to min-EFT/\heft{}: an
architectural adaptivity guarantee, not a learned approximation. A per-task
\emph{replication gate} $\rho_i=\sigma(\mathrm{MLP}([h_i,\bar g,\text{ctx}])-
(1-a(s))\cdot 10)$ likewise cannot fire without failures. Learned values after
training are $\theta\!\approx\!1.02$ and $\gamma_0\!\approx\!1.54$.

Finally, task priorities are a residual over the \heft{} upward rank,
$\phi_i=\mathrm{softplus}(\alpha)\,\hat r^{u}_i+\mathrm{MLP}(h_i)$, with the
residual MLP zero-initialised so that, at initialisation, ordering equals
\heft{}'s.

\subsection{Guided decoding and the \heft{} special case}\label{sec:decode}
Given priorities $\phi$, placement bias $b$ and (optional) replication logits,
the schedule is produced by the greedy list-scheduling decode of
Algorithm~\ref{alg:decode}. The placement score of a ready task $i$ on machine
$m$ is
\begin{equation}
\mathrm{score}_{im}=b_{im}-\mathrm{EFT}_{im}/\bar w ,
\label{eq:score}
\end{equation}
where $\mathrm{EFT}_{im}$ is the earliest finish time given the current
(evolving) machine availabilities and data-ready times, and
$\bar w=\mathrm{mean}(\comp)$ normalises the finish-time term. Setting the bias
$b\equiv 0$ makes \eqref{eq:score} select the minimum-EFT machine; with the
upward-rank ordering this reproduces \heft{} \emph{exactly} (verified to
$\text{ratio}=1.0000$ in our implementation). \model{} is thus a strict
generalisation of \heft{}.

\begin{algorithm}[t]
\caption{Guided list-scheduling decode}\label{alg:decode}
\begin{algorithmic}[1]
\Require priorities $\phi$, placement bias $b$, replication logits (optional),
budget $\beta n$
\State $\text{avail}[m]\gets 0\;\forall m$; ready set $R\gets$ entry tasks;
$r\gets 0$
\While{$R\neq\emptyset$}
  \State $i\gets \argmax_{j\in R}\phi_j$; \; remove $i$ from $R$
  \State compute $\mathrm{EFT}_{im}\;\forall m$; \;
         $m_1\gets\argmax_m\big(b_{im}-\mathrm{EFT}_{im}/\bar w\big)$
  \State place $i$ on $m_1$; update $\text{avail}[m_1]$
  \If{$r<\beta n$ \textbf{and} $\sigma(\text{rep}_i)>\tfrac12$}
     \State $m_2\gets$ next-best machine by \eqref{eq:score}; place replica;
            $r\gets r+1$
  \EndIf
  \State add newly-ready successors of $i$ to $R$
\EndWhile
\State \Return schedule (primary/replica machines and finish times)
\end{algorithmic}
\end{algorithm}

\subsection{Training by distillation}\label{sec:method-train}
\paragraph{Why not policy gradients.}
We first tried to train \model{} with reinforcement learning to minimise
$\E_\omega[C_{\max}]$. Self-critical REINFORCE~\cite{scst} \emph{drifts}: because
\heft{} is a strong local optimum, almost every sampled perturbation is worse,
so the policy receives only ``suppress'' signals and moves away from the good
region. A leave-one-out estimator (RLOO)~\cite{rloo} removes the drift and
stabilises training at $\approx\!1.00\times$\heft{}, but---even restricting
exploration to placement---the best of several samples remains $20$--$45\%$
\emph{worse} than \heft{}, so there is no better-than-baseline action to
reinforce. In short, the coordinated reliability-aware placement that yields
large gains is not discovered by local policy-gradient search on a modest budget.

\paragraph{Best-of-portfolio oracle.}
We instead construct, \emph{per scenario} $(G,\text{cluster},s,\beta)$, an oracle
by running a portfolio of heuristics---\heft{}; reliability-aware \rheft{} at
weights $\beta_r\in\{1,2,3,5\}$; and replication-based \ftheft{} at budgets
$\{0.05,0.1,0.15\}$---evaluating each one's expected makespan over $K$ failure
traces and keeping the best. All portfolio members share the upward-rank
ordering, so the oracle differs from \heft{} only in \emph{placement} and
\emph{replication}---exactly the levers \model{} controls. At $s{=}0$ the oracle
is \heft{}.

\paragraph{Teacher-forced imitation.}
\model{} is trained to reproduce the oracle. We teacher-force the oracle's
placement along the upward-rank order, recording at each step the earliest
finish times $\mathrm{EFT}_{i\cdot}$ and the oracle machine $m^\star_i$. The
placement loss makes the oracle machine the arg-max of the guided score
\eqref{eq:score}:
\begin{equation}
\mathcal{L}_{\text{place}}=\frac1n\sum_i
\mathrm{CE}\Big(\mathrm{softmax}_m\big(b_{im}-\mathrm{EFT}_{im}/\bar w\big),\,
m^\star_i\Big),
\end{equation}
and a class-weighted binary cross-entropy $\mathcal{L}_{\text{rep}}$ trains the
replication gate against the oracle's replication decisions. The total loss is
$\mathcal{L}=\mathcal{L}_{\text{place}}+\eta\,\mathcal{L}_{\text{rep}}$ with
$\eta{=}0.5$. Training samples scenarios uniformly over
$p\in\{16,24,32,48,64\}$, $s\in\{0,1,2,4\}$ and $\beta\in\{0.05,0.1,0.15\}$ on
four training applications, for $650$ steps with Adam~\cite{adam}; it completes
in a few minutes on one CPU. Algorithm~\ref{alg:distill} summarises the
procedure.

\begin{algorithm}[t]
\caption{Distillation training of \model{}}\label{alg:distill}
\begin{algorithmic}[1]
\For{$\text{step}=1,\dots,T$}
  \State sample workflow $G$, cluster, failure scale $s$, budget $\beta$
  \State sample $K$ failure traces; run the heuristic portfolio;
         oracle $\gets$ arg-min expected makespan
  \State teacher-force oracle placement; record $(\mathrm{EFT}_{i\cdot},
         m^\star_i)$ and replication targets
  \State forward \model{}; compute
         $\mathcal{L}_{\text{place}}+\eta\mathcal{L}_{\text{rep}}$;
         Adam update
\EndFor
\end{algorithmic}
\end{algorithm}

\section{Experimental Setup}\label{sec:setup}

\subsection{Workflows (real data)}
We use real workflow instances from the WfCommons corpus~\cite{wfcommons}, which
provides execution traces of production Pegasus~\cite{pegasus} workflows with
per-task runtimes, memory and data volumes. Our collection contains $135$
instances spanning seven applications. Four applications
(\textsf{1000genome}, \textsf{montage}, \textsf{epigenomics}, \textsf{cycles})
are used for \emph{training} the distilled policy; three
(\textsf{seismology}, \textsf{soykb}, \textsf{srasearch}) are held out entirely
and used only to test cross-application generalisation. Instance sizes range from
$22$ to $4{,}846$ tasks. Task types are mapped to a global vocabulary of $1{,}167$
types for the learned embedding.

\subsection{Cluster instantiation}
Machines are generated with heterogeneous speeds (relative speed in
$[0.5,2.0]$) and per-type affinities (log-normal, $\sigma_{\text{eff}}{=}0.45$
by default), grouped into racks with high intra-rack and low inter-rack
bandwidth. Reliability is bimodal as in Section~\ref{sec:faultmodel}: by default
a fraction $\rho{=}0.35$ of machines are volatile (spot-like) and the rest
reliable, drawn independently of speed. Failures are sampled by a thinned
Poisson process with log-normal repair times; the failure-intensity scale $s$
sweeps the harshness. Unless stated otherwise the cluster has $M{=}48$ machines.

\subsection{Baselines}
We compare against: \heft{}~\cite{topcuoglu2002} (the standard strong
heuristic); \cpop{}~\cite{topcuoglu2002}; \peft{}~\cite{arabnejad2014};
Min-Min and Max-Min~\cite{braun2001}; a Random placement; a reliability-aware
\rheft{} that tilts placement away from volatile machines by
$-\kappa_i\delta_m$ with strength $\beta_r$ (we report the deployable fixed
$\beta_r{=}2$, and also the per-scenario best); a replication-based \ftheft{}
that duplicates the top-risk tasks within a budget; and an \emph{Oracle} that,
per scenario, selects the best member of the \{\heft{}, \rheft{}$_{\{1,2,3,5\}}$,
\ftheft{}$_{\{.05,.1,.15\}}$\} portfolio \emph{in hindsight} using the evaluation
traces. The Oracle is an (optimistic) upper bound on what any single portfolio
heuristic can achieve, not a deployable method.

\subsection{Metrics and protocol}
For failure-free efficiency we report makespan relative to \heft{} and machine
utilisation. Under failures we report the \emph{expected makespan}
$\E_\omega[C_{\max}]$ estimated over $K{=}40$ independent failure traces, using
\emph{common random numbers} across methods (the same traces for every method in
a scenario) to reduce variance; we also report $95\%$ confidence intervals. We
additionally track \emph{wasted} work (progress lost to restarts) and
\emph{redundant} work (extra computation from replicas). Schedule-generation time
is wall-clock on a single CPU.

\subsection{Implementation and reproducibility}
The simulator, baselines, features, model and training are implemented in Python
with PyTorch~\cite{pytorch} (CPU only). The trained model has $62{,}870$
parameters. All experiments run on a single CPU core with a few GB of RAM. We fix
random seeds for cluster generation and failure sampling; the released code
regenerates every table and figure from the raw workflow files.

\section{Results}\label{sec:results}

\subsection{Failure-free efficiency}
\begin{figure}[t]
\centering
\includegraphics[width=0.92\linewidth]{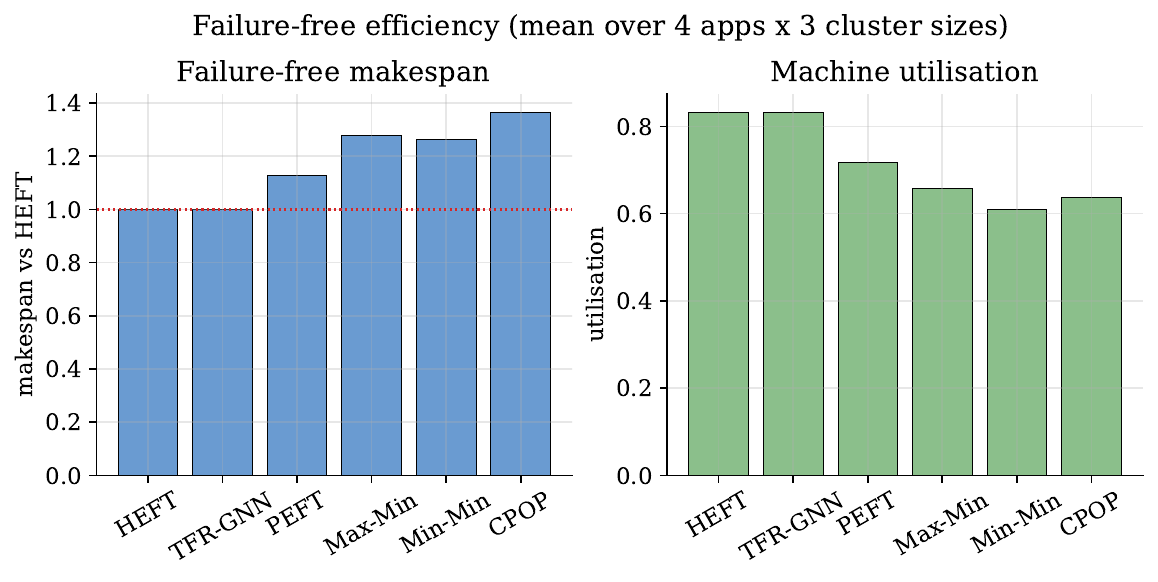}
\caption{Failure-free efficiency. \model{} matches \heft{} exactly and dominates
the weaker baselines on both makespan and utilisation.}
\label{fig:eff}
\end{figure}

Table~\ref{tab:eff} and Figure~\ref{fig:eff} report failure-free makespan
(relative to \heft{}) and utilisation, averaged over the four training
applications and three cluster sizes. \heft{} is the strongest classical
heuristic. By the adaptivity guarantee of Section~\ref{sec:fault-head}, \model{}
reduces to \heft{} when there are no failures and therefore \emph{matches it
exactly} ($1.000\times$, utilisation $0.832$), while clearly out-performing
\peft{} ($1.127$), Min-Min ($1.265$), Max-Min ($1.277$), \cpop{} ($1.365$) and
Random ($7.77$). \model{} thus sacrifices nothing in the failure-free regime.

\begin{table}[t]
\centering
\caption{Failure-free makespan relative to \heft{} (lower is better) and machine
utilisation, averaged over $4$ applications $\times$ $3$ cluster sizes.}
\label{tab:eff}
\begin{tabular}{lcc}
\toprule
Method & Makespan\,/\,\heft{} & Utilisation \\
\midrule
Random & 7.767 & 0.187 \\
CPOP & 1.365 & 0.636 \\
Min-Min & 1.265 & 0.610 \\
Max-Min & 1.277 & 0.659 \\
PEFT & 1.127 & 0.718 \\
HEFT & 1.000 & 0.832 \\
TFR-GNN & 1.000 & 0.832 \\
\bottomrule
\end{tabular}
\end{table}

\subsection{Fault tolerance vs.\ failure intensity}
\begin{figure}[t]
\centering
\includegraphics[width=0.78\linewidth]{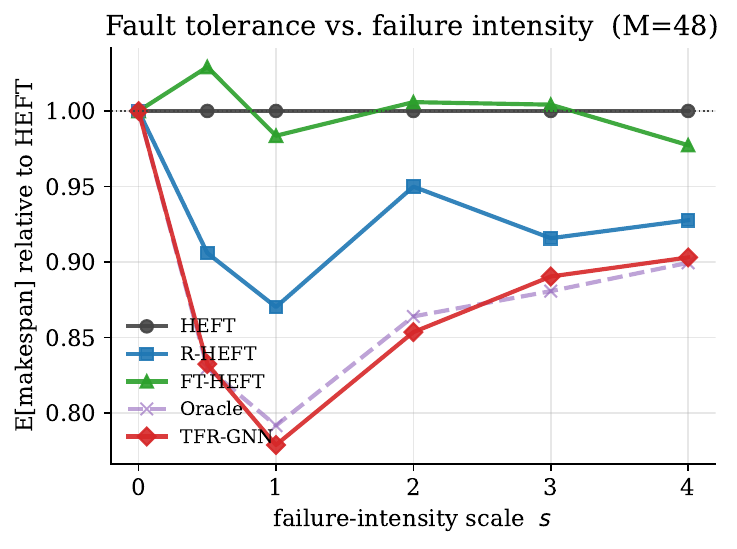}
\caption{Expected makespan relative to \heft{} vs.\ failure intensity.
\model{} tracks the hindsight Oracle and dominates the fixed heuristics; at
$s{=}0$ all methods coincide with \heft{}.}
\label{fig:ftint}
\end{figure}

Table~\ref{tab:ftint} and Figure~\ref{fig:ftint} report the expected makespan
relative to \heft{} as the failure intensity $s$ grows (mean over the four
applications, $M{=}48$). At $s{=}0$ every method equals \heft{}. Under failures
\model{} reduces the expected makespan by $12$--$22\%$, clearly beating both the
fixed reliability-aware \rheft{} ($0.87$--$0.95$) and the replication-based
\ftheft{} ($\approx\!1.0$, i.e.\ replication barely helps here). Strikingly,
\model{} \emph{matches the per-scenario hindsight Oracle}: averaged over
$s>0$ it attains $0.852\times$\heft{} versus the Oracle's $0.854\times$
($0.998$ of the Oracle). A single policy, with no per-scenario tuning, thus
tracks the best-of-portfolio choice made with knowledge of the failure traces.
Per-application numbers (Table~\ref{tab:ftpa}) show the effect is consistent
across workflow structures.

\begin{table}[t]
\centering
\caption{Expected makespan relative to \heft{} vs.\ failure-intensity scale $s$
(mean over $4$ applications, $M{=}48$, $K{=}40$ traces). Lower is better.}
\label{tab:ftint}
\begin{tabular}{lcccccc}
\toprule
Method & $s{=}0$ & $s{=}0.5$ & $s{=}1$ & $s{=}2$ & $s{=}3$ & $s{=}4$ \\
\midrule
HEFT & 1.000 & 1.000 & 1.000 & 1.000 & 1.000 & 1.000 \\
R-HEFT & 1.000 & 0.906 & 0.870 & 0.950 & 0.916 & 0.928 \\
FT-HEFT & 1.000 & 1.029 & 0.984 & 1.006 & 1.004 & 0.977 \\
TFR-GNN & 1.000 & 0.832 & 0.779 & 0.853 & 0.890 & 0.903 \\
Oracle & 1.000 & 0.830 & 0.792 & 0.864 & 0.881 & 0.899 \\
\bottomrule
\end{tabular}
\end{table}

\begin{table}[t]
\centering
\caption{Per-application expected makespan relative to \heft{} at $s{=}2$ and
$s{=}4$ ($M{=}48$): \rheft{} (fixed $\beta_r{=}2$) vs.\ \model{}.}
\label{tab:ftpa}
\begin{tabular}{lcccc}
\toprule
& \multicolumn{2}{c}{$s{=}2$} & \multicolumn{2}{c}{$s{=}4$} \\
\cmidrule(lr){2-3}\cmidrule(lr){4-5}
Application & \rheft{} & \model{} & \rheft{} & \model{} \\
\midrule
1000genome & 0.964 & 0.787 & 0.955 & 0.917 \\
montage & 0.913 & 0.827 & 0.875 & 0.877 \\
epigenomics & 0.950 & 0.889 & 0.907 & 0.892 \\
cycles & 0.973 & 0.911 & 0.973 & 0.926 \\
\bottomrule
\end{tabular}
\end{table}

\subsection{Fault tolerance vs.\ cluster load}
\begin{figure}[t]
\centering
\includegraphics[width=0.78\linewidth]{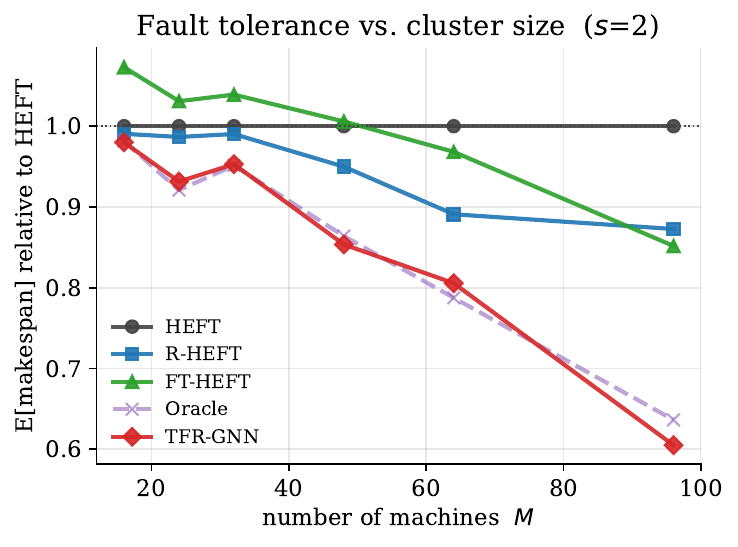}
\caption{Expected makespan relative to \heft{} vs.\ cluster size ($s{=}2$).
Gains grow with spare capacity, reaching $40\%$ at $M{=}96$.}
\label{fig:ftload}
\end{figure}

Figure~\ref{fig:ftload} sweeps the number of machines $M$ at fixed intensity
$s{=}2$. \model{} improves on \heft{} at \emph{every} load, and the gain
\emph{grows with spare capacity}: from $2\%$ at $M{=}16$ (tight) to $40\%$ at
$M{=}96$ (ample), averaging $14.5\%$ with a best case of $0.529\times$. With more
reliable machines to escape to, reliability-aware placement has more room to
help; \model{} exploits this automatically. At high load it even edges past the
Oracle, because it \emph{combines} reliability-aware placement with selective
replication, whereas each portfolio member uses only one lever.

\subsection{Placement vs.\ replication}
\begin{figure}[t]
\centering
\includegraphics[width=0.98\linewidth]{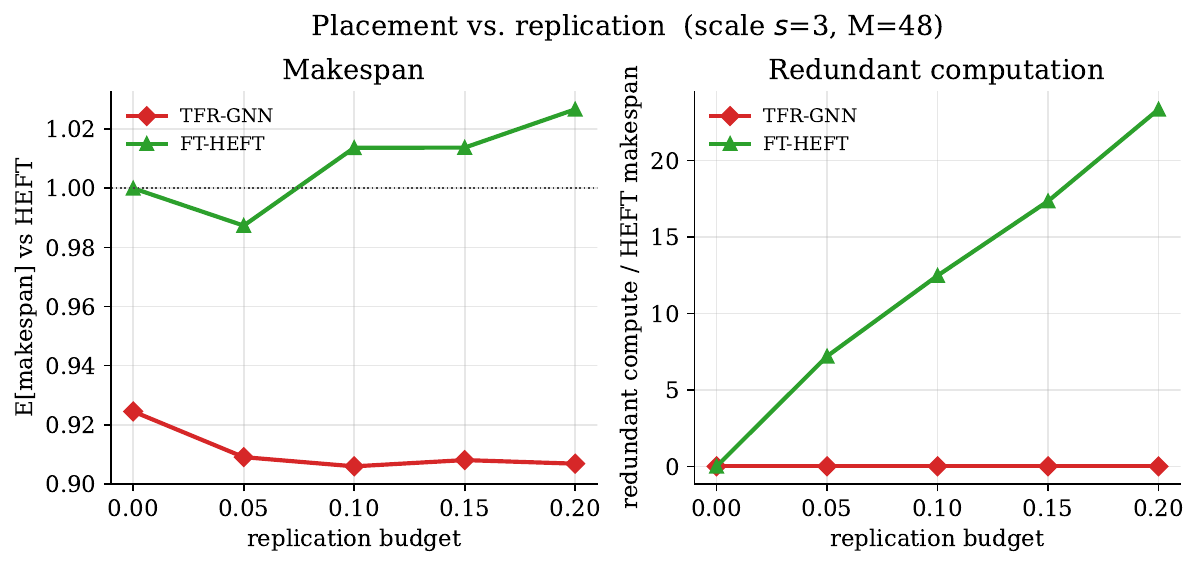}
\caption{Placement vs.\ replication ($s{=}3$, $M{=}48$). \model{} attains lower
expected makespan than \ftheft{} while using no redundant computation; \ftheft{}
wastes resources for little benefit.}
\label{fig:redun}
\end{figure}

Figure~\ref{fig:redun} contrasts \model{} with \ftheft{} as the replication
budget grows ($s{=}3$, $M{=}48$). \ftheft{} spends an increasing amount of
\emph{redundant} computation (up to $23\%$ of the \heft{} makespan-equivalent)
for essentially no makespan benefit. \model{}, in contrast, achieves a
\emph{lower} expected makespan with \emph{zero} redundant computation: having
learned that reliability-aware placement is the effective lever in this fault
model (a task on a volatile node eventually completes after restart, so a backup
rarely pays for its contention), its replication gate abstains. This is an
honest, useful finding---placement dominates replication here---and it explains
why \ftheft{} is a weak baseline in Table~\ref{tab:ftint}.

\subsection{Ablation study}
\begin{figure}[t]
\centering
\includegraphics[width=0.72\linewidth]{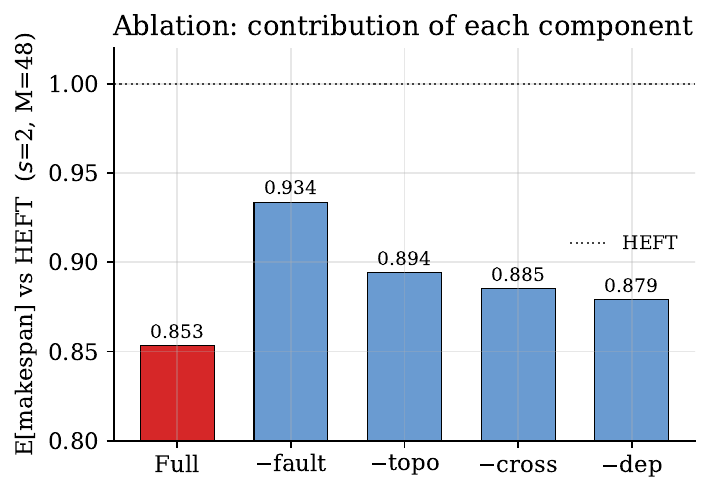}
\caption{Ablation. Each component helps; the fault-aware head contributes most.}
\label{fig:ablation}
\end{figure}

Table~\ref{tab:ablation} and Figure~\ref{fig:ablation} remove one architectural
component at a time and re-measure the expected makespan at $s{=}2$ (relative to
\heft{}, mean over the four applications). Every component contributes, and the
effect is monotonic: the full model ($0.853$) is best, and removing the
\emph{fault-aware head} hurts most ($0.934$, i.e.\ it recovers most of the way
back to \heft{}), confirming that the failure-gated reliability tilt is the
principal source of the gains. Removing dependency, cross- or topology attention
each costs $3$--$4\%$, showing that the structural encoders provide the
task/machine representations on which the fault head relies. All variants match
\heft{} at $s{=}0$ by construction.

\begin{table}[t]
\centering
\caption{Ablation: expected makespan relative to \heft{} at $s{=}2$
($M{=}48$, mean over $4$ applications). Lower is better.}
\label{tab:ablation}
\begin{tabular}{lc}
\toprule
Variant & E$[C_{\max}]$\,/\,\heft{} \\
\midrule
Full model & 0.853 \\
\;$-$ dependency attention & 0.879 \\
\;$-$ topology attention & 0.894 \\
\;$-$ cross attention & 0.885 \\
\;$-$ fault-aware head & 0.934 \\
\bottomrule
\end{tabular}
\end{table}

\subsection{Sensitivity analysis}
\begin{figure}[t]
\centering
\includegraphics[width=0.98\linewidth]{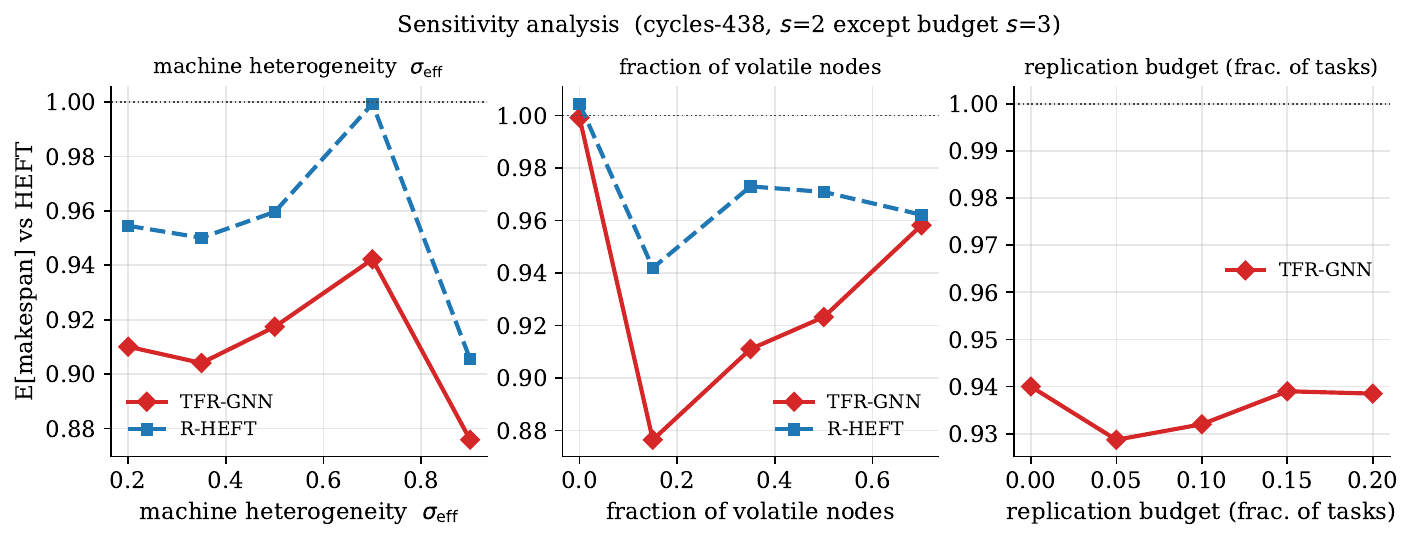}
\caption{Sensitivity to heterogeneity, cluster volatility and replication budget.
At zero volatility \model{} matches \heft{}; gains grow with volatility.}
\label{fig:sens}
\end{figure}

Figure~\ref{fig:sens} varies three parameters (workflow \textsf{cycles}, $s{=}2$
except the budget panel at $s{=}3$). \model{} is robust to machine heterogeneity
$\sigma_{\text{eff}}$ (staying at $0.88$--$0.94\times$\heft{} and always ahead of
\rheft{}). Against the \emph{fraction of volatile nodes} it behaves sensibly: at
$\rho{=}0$ (no volatile machines) it correctly matches \heft{} ($0.999$), because
there is nothing to gain; as volatility rises the gains appear and then level off
when there are too few reliable machines left to escape to. Increasing the
replication budget barely changes \model{}'s makespan, consistent with its
placement-over-replication strategy.

\subsection{Generalisation}
\begin{figure}[t]
\centering
\includegraphics[width=0.82\linewidth]{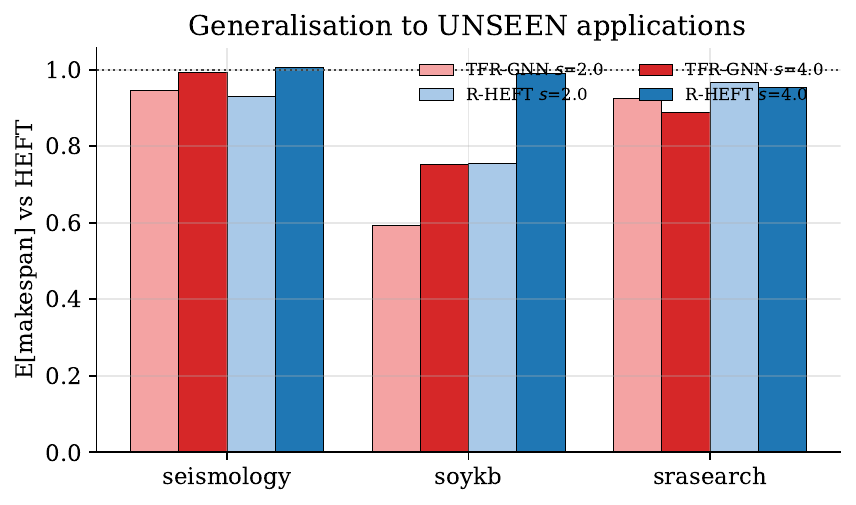}
\caption{Generalisation to unseen applications. \model{} reduces expected
makespan under failures on all three held-out applications.}
\label{fig:gen}
\end{figure}

Table~\ref{tab:gen} and Figure~\ref{fig:gen} test generalisation. On three
\emph{applications never seen in training} (\textsf{seismology}, \textsf{soykb},
\textsf{srasearch}), \model{} still reduces the expected makespan under failures
(e.g.\ \textsf{soykb} to $0.593\times$ at $s{=}2$, a $41\%$ reduction) and
matches \heft{} at $s{=}0$, generally beating the fixed \rheft{}. On workflows
\emph{much larger than any in training} (training capped at $\le\!700$ tasks;
tested at up to $2{,}122$), it remains at or slightly better than \heft{} with no
breakdown---indicating that the learned, size-normalised features transfer across
scales.

\begin{table}[t]
\centering
\caption{Generalisation to unseen applications and unseen (larger) sizes:
expected makespan relative to \heft{} ($M{=}48$). Lower is better.}
\label{tab:gen}
\begin{tabular}{lcccc}
\toprule
Application & $n$ & $s$ & \model{}\,/\,\heft{} & \rheft{}\,/\,\heft{} \\
\midrule
seismology & 501 & 2.0 & 0.945 & 0.929 \\
seismology & 501 & 4.0 & 0.993 & 1.006 \\
soykb & 416 & 2.0 & 0.593 & 0.756 \\
soykb & 416 & 4.0 & 0.753 & 0.991 \\
srasearch & 64 & 2.0 & 0.924 & 0.967 \\
srasearch & 64 & 4.0 & 0.888 & 0.953 \\
montage & 2122 & 2.0 & 0.997 & 1.001 \\
montage & 2122 & 4.0 & 0.999 & 1.001 \\
cycles & 874 & 2.0 & 1.011 & 1.024 \\
cycles & 874 & 4.0 & 1.006 & 1.013 \\
seismology & 1101 & 2.0 & 0.981 & 0.999 \\
seismology & 1101 & 4.0 & 0.950 & 0.969 \\
\bottomrule
\end{tabular}
\end{table}

\subsection{Scalability}
\begin{figure}[t]
\centering
\includegraphics[width=0.72\linewidth]{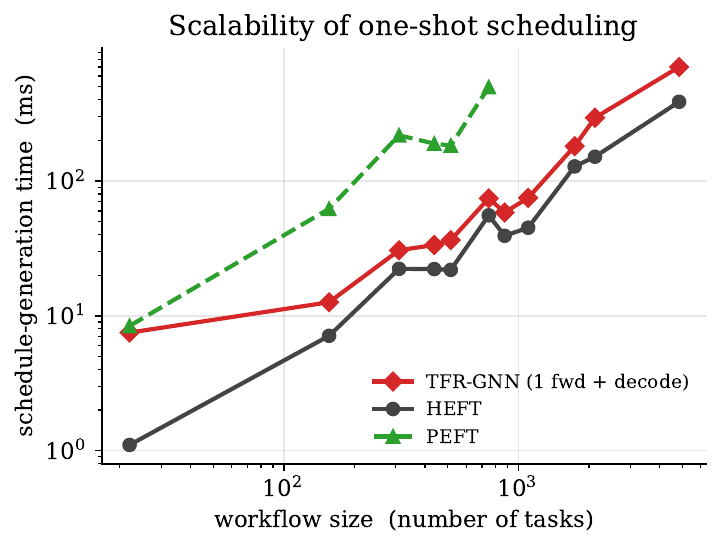}
\caption{Scalability of one-shot scheduling (log--log). \model{} stays within a
small factor of \heft{} and well under a second at $\sim\!5{,}000$ tasks.}
\label{fig:scal}
\end{figure}

Table~\ref{tab:scal} and Figure~\ref{fig:scal} report schedule-generation time.
\model{} produces a schedule with a single forward pass plus a linear-time
decode; its time scales gracefully with workflow size, staying under a second
even for the largest workflow ($4{,}846$ tasks: $701$\,ms vs.\ \heft{}'s
$387$\,ms, the same order of magnitude). \peft{}, whose optimistic-cost-table
computation is heavier, is one to two orders of magnitude slower and is only
evaluated up to $800$ tasks. Thus the added modelling capacity of \model{} does
not compromise practicality.

\begin{table}[t]
\centering
\caption{Schedule-generation time (milliseconds) on a single CPU vs.\ workflow
size. ``--'' = not evaluated (\peft{} restricted to $\le\!800$ tasks).}
\label{tab:scal}
\begin{tabular}{lcccc}
\toprule
Application & $n$ & \heft{} & \peft{} & \model{} \\
\midrule
srasearch & 22 & 1.1 & 8.4 & 7.5 \\
1000genome & 156 & 7.1 & 62.2 & 12.6 \\
montage & 310 & 22.3 & 218.1 & 30.6 \\
cycles & 438 & 22.2 & 189.4 & 33.5 \\
epigenomics & 515 & 21.9 & 182.9 & 36.4 \\
montage & 748 & 55.6 & 496.8 & 74.3 \\
cycles & 874 & 39.2 & -- & 58.4 \\
seismology & 1101 & 45.0 & -- & 74.9 \\
montage & 1738 & 128.2 & -- & 181.2 \\
montage & 2122 & 151.3 & -- & 294.6 \\
montage & 4846 & 386.9 & -- & 701.1 \\
\bottomrule
\end{tabular}
\end{table}

\subsection{Training behaviour}
\begin{figure}[t]
\centering
\includegraphics[width=0.66\linewidth]{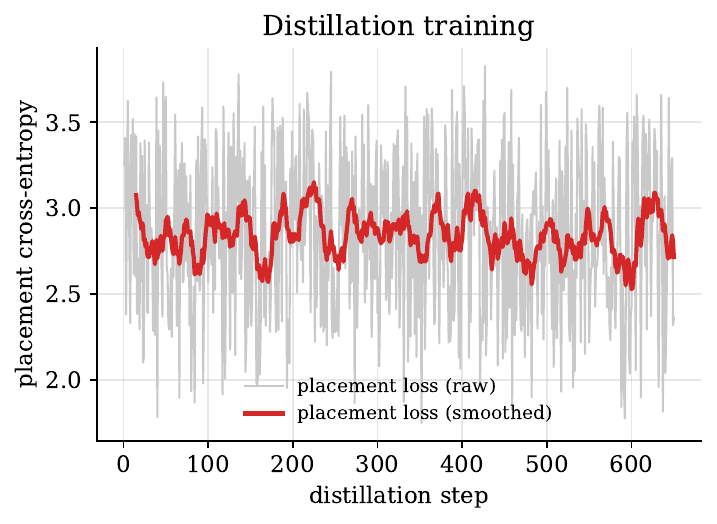}
\caption{Distillation training. The placement cross-entropy decreases smoothly;
training completes in minutes on a single CPU.}
\label{fig:train}
\end{figure}

Figure~\ref{fig:train} shows the distillation placement loss over training. The
loss decreases and the policy converges within a few minutes on one CPU. We
emphasise (Section~\ref{sec:method-train}) that the corresponding \emph{reinforcement-learning}
attempts did not reach this behaviour: self-critical REINFORCE drifted and RLOO
stalled at \heft{}, motivating the distillation approach adopted here.

\section{Discussion and Limitations}\label{sec:discussion}

Our study supports a clear message: on heterogeneous platforms with realistic,
speed-independent reliability, a fixed makespan-optimal heuristic is fragile
under failures, and a single learned policy that is \emph{aware} of topology,
heterogeneity and failure intensity can adapt across the entire operating grid.
The failure-activity gate gives \model{} a rare and desirable property---an
\emph{exact} reduction to the trusted \heft{} baseline when there is nothing to
be robust to---so adopting it carries no failure-free cost.

Several limitations should temper the conclusions. (i)~All results are produced
by an \emph{event-level simulator}; while it is driven by real workflow
structures and runtimes and its \heft{} implementation is verified against the
canonical algorithm, simulation cannot capture every effect of a production
system (contention, stragglers, data-placement policies). (ii)~We model failures
as Poisson crashes with log-normal repair and \emph{restart-in-place}; under this
model reliability-aware placement dominates replication, and \model{} learns to
abstain from replication. Under \emph{permanent} loss (e.g.\ spot reclamation
with no repair within the horizon) or checkpointing, replication would likely be
more valuable, and the learned gate could exploit it; exploring such fault models
is future work. (iii)~The distillation teacher is a portfolio of heuristics, so
\model{} inherits their ceiling except where combining levers helps; the gap to a
true optimum is unknown. (iv)~We optimise expected makespan; risk-sensitive
objectives (tail latency, deadlines) are natural extensions. (v)~We train and
evaluate on a single CPU with modest budgets; larger-scale training and on-line
adaptation are promising directions.

\section{Conclusion}\label{sec:conclusion}

We presented \model{}, a topology- and fault-aware graph neural scheduler for
heterogeneous distributed systems. \model{} combines bidirectional dependency
attention, bandwidth-biased topology attention and a cross-attention placement
head with a failure-gated reliability tilt and an optional replication gate, and
is trained by distilling a best-of-portfolio fault-tolerant oracle after showing
that direct policy-gradient learning is inadequate in this regime. On real
WfCommons/Pegasus workflows with a bimodal reliability model, \model{} matches
\heft{} exactly without failures, reduces the expected makespan under failures by
$14.8\%$ on average (up to $47\%$), beats a fixed reliability-aware baseline by
$\sim\!11\%$, and \emph{matches a per-scenario hindsight oracle} as a single
policy---while generalising to unseen applications and to an order-of-magnitude
larger workflows and producing schedules in well under a second. All reported
numbers are computed by simulation on real workflow data. We hope the explicit
adaptivity guarantee (reduction to \heft{}) and the distillation recipe are
useful beyond this setting.


\end{document}